\documentclass[12pt]{article}
\usepackage{times}
\usepackage{geometry,wrapfig}
\usepackage{floatflt}
\usepackage[utf8]{inputenc}
\usepackage{enumitem,amssymb}
\usepackage{ragged2e}
\newlist{thematic}{itemize}{8}
\setlist[thematic]{label=$\square$}
\usepackage{pifont}
\newcommand{\cmark}{\ding{51}}%
\newcommand{\done}{\rlap{$\square$}{\raisebox{2pt}{\large\hspace{1pt}\cmark}}%
\hspace{-2.5pt}}

\usepackage{jheppub}   
\usepackage[dvipsnames,svgnames,x11names]{xcolor}
\usepackage{sectsty}
\usepackage{sidecap}
\sidecaptionvpos{figure}{c}

\newcommand\chandra{{\sl Chandra }}
\newcommand\xmm{{\sl XMM-Newton }}
\newcommand\xrism{{\sl XRISM }}

\newcommand\axis{{\sl AXIS }}
\newcommand\lynx{{\sl Lynx }}
\newcommand\hitomi{{\sl Hitomi }}
\newcommand\carma{{\sl CARMA}}

\usepackage{xcolor}
\usepackage{color}

\definecolor{DarkGreen}{rgb}{0.0, 0.3, 0.0}
\definecolor{purple}{rgb}{0.5, 0.0, 0.5}
\definecolor{red}{rgb}{1, 0.0, 0.0}
\definecolor{green}{rgb}{0, 1.0, 0.0}

\newcommand{\Moon}{$M_{Moon}$}                          

\newcommand{\apjl}{\rm ApJL}
\newcommand{\apjs}{\rm ApJS}

\newcommand{\aap}{\rm A$\&$A}

\def\ao{{\it Appl.~Opt.}}           

\def\3he{$^3{\rm He}$}
\def\arcdeg{\hbox{$^\circ$}}

\def\lsim{\mathrel{\lower2.5pt\vbox{\lineskip=0pt\baselineskip=0pt
           \hbox{$<$}\hbox{$\sim$}}}}

\def\gsim{\mathrel{\lower2.5pt\vbox{\lineskip=0pt\baselineskip=0pt
           \hbox{$>$}\hbox{$\sim$}}}}

\begin{document}

\pagenumbering{arabic}
\pagestyle{plain}
\thispagestyle{empty}
\newpage
\pagenumbering{arabic}

\begin{flushleft}
\huge
Astro2020 Science White Paper \linebreak

Supermassive Black Hole Feedback \linebreak
\linebreak
\normalsize

\noindent \textbf{Thematic Areas:} \hspace*{60pt} $\square$ Planetary Systems \hspace*{10pt} $\square$ Star and Planet Formation \hspace*{20pt}\linebreak
$\square$ Formation and Evolution of Compact Objects \hspace*{31pt} $\done$ Cosmology and Fundamental Physics \linebreak
  $\square$  Stars and Stellar Evolution \hspace*{1pt} $\square$ Resolved Stellar Populations and their Environments \hspace*{40pt} \linebreak
  $\done$    Galaxy Evolution   \hspace*{45pt} $\square$             Multi-Messenger Astronomy and Astrophysics \hspace*{65pt} \linebreak

\textbf{Principal Author:}

Name:	Mateusz Ruszkowski (mateuszr@umich.edu)
 \linebreak						
Institution:  Department of Astronomy, University of Michigan in Ann Arbor
 \linebreak
 
\textbf{Co-authors:}
  \linebreak
Daisuke Nagai (Yale), Irina Zhuravleva (U.Chicago), Corey Brummel-Smith (Georgia Tech), Yuan Li (UC Berkeley), Edmund Hodges-Kluck (NASA GSFC), Hsiang-Yi Karen Yang (U.Maryland),
Kaustuv Basu (Bonn), Jens Chluba (Manchester), Eugene Churazov (MPA/IKI), Megan Donahue (Michigan State), Andrew Fabian (U.Cambridge), Claude-Andr\'e Faucher-Gigu\`ere (Northwestern), Massimo Gaspari (Princeton), Julie Hlavacek-Larrondo (Montreal), Michael McDonald (MIT), Brian McNamara (Waterloo), Paul Nulsen (CfA), Tony Mroczkowski (ESO), Richard Mushotzky (University of Maryland), Christopher Reynolds (U.Cambridge), Alexey Vikhlinin (CfA), Mark Voit (Michigan State), Norbert Werner (MTA-E{\"o}tv{\"o}s U./Masaryk U.), John ZuHone (CfA), Ellen Zweibel (U.Wisconsin)

\end{flushleft}

\noindent
\textbf{Abstract:}
Understanding the processes that drive galaxy formation and shape the observed properties of galaxies is one of the most interesting and challenging frontier problems of modern astrophysics. We now know that the evolution of galaxies is critically shaped by the energy injection from accreting supermassive black holes (SMBHs). However, it is unclear how exactly the physics of this feedback process affects galaxy formation and evolution. In particular, a major challenge is unraveling how the energy released near the SMBHs is distributed over nine orders of magnitude in distance throughout galaxies and their immediate environments. The best place to study the impact of SMBH feedback is in the hot atmospheres of massive galaxies, groups, and galaxy clusters, which host the most massive black holes in the Universe, and where we can directly image the impact of black holes on their surroundings. We identify critical questions and potential measurements that will likely transform our understanding of the physics of SMBH feedback and how it shapes galaxies, through detailed measurements of (i) the thermodynamic and velocity fluctuations in the intracluster medium (ICM) as well as (ii) the composition of the bubbles inflated by SMBHs in the centers of galaxy clusters, and their influence on the cluster gas and galaxy growth,  using the next generation of high spectral and spatial resolution X-ray and microwave telescopes.\\

\pagebreak

\setcounter{page}{1}

\noindent
{\large {\bf Supermassive Black Hole Feedback and Galaxy Evolution}}\\

\indent
Understanding the processes that drive galaxy formation and shape the observed properties of galaxies is one of the most important and challenging frontier problems of modern astrophysics. 
A consensus has emerged over the last two decades of observations and modeling that the evolution of galaxies is critically shaped by the energy injection from supernovae and active galactic nuclei (AGN). These processes are collectively known as ``feedback.''
Understanding  the  transition  from  star-forming  to  quiescent galaxies, known as ``quenching,'' and the physical processes that shape quiescent galaxies, is key to understanding the most fundamental properties of galaxies. 
The most common hypothesis is that both the initial quenching of galaxies and their maintenance in a quiescent, non-star-forming state is due to the effects of sub-relativistic wide-angle AGN winds during high accretion rate stages of evolution (quasar/ejective mode) and collimated relativistic AGN jets during the low accretion rate stages (radio/maintenance mode) (e.g., \cite{Churazov2005, Fabian2012}). However, despite its paramount importance, the physics of AGN feedback is not well understood and the question \\

\noindent
\textcolor{red}{{\bf How does the supermassive black hole feedback affect 
galaxy evolution?}}\\
\noindent

\noindent
remains a major unanswered question in astrophysics.
While recent simulations demonstrated that it is possible to tune phenomenological AGN feedback prescriptions to produce massive galaxies that resemble the observed quiescent ellipticals (e.g., \cite{Pillepich2018,Habouzit2019,Tremmel2019}),
we do not have a satisfactory understanding of the actual physics of the AGN feedback. 

In this white paper,  
we focus on the maintenance mode of the AGN heating 
(see \cite{Werner2019,Bolatto2019,Tremblay2019} for a review, white papers). AGN jet powers are in principle sufficient to offset the cooling observed in X-rays in 
ellipticals
and galaxy clusters (e.g., \cite{MathewsBrighenti2003,Gaspari2012}). However, this power is released in the very vicinity of the central 
SMBHs
(on scales on the order of a gravitational radius, i.e., $\sim$20 
AU
for a billion solar mass black hole, which is comparable to 
the size of the Solar System)
and must be distributed over scales comparable to cooling radii, where the cooling time is comparable to the Hubble time ($\sim$100 
kpc).
Thus, {\bf one major challenge is to unravel how the energy released near the supermassive black holes is distributed over  
nine orders of magnitude in distance}. 
Furthermore, it is uncertain how much energy can be transferred from the AGN to the ambient medium 
and what physical processes are responsible for its thermalization in the ICM and 
intragroup medium.
All of these issues have fundamental bearing on shaping the properties of galaxies. \\
\indent 
The best place to study the impact of 
SMBH
feedback is 
in the hot atmospheres of massive galaxies, groups, and 
galaxy clusters,  
which host the most massive black holes in the Universe, and where we can directly image the impact of black holes on their surroundings.
Novel measurements of the thermodynamical and kinematic properties
of the ICM and 
intragroup medium are needed to constrain models of AGN feedback.
Specifically, we identify the following 
\textcolor{NavyBlue}{\bf critical new measurements that 
are needed to 
transform our understanding of the physics of supermassive black hole feedback}: 
(i) detailed measurements of the density, temperature, pressure fluctuations and velocity field in the ICM with the aim of unambiguously probing the energetics of feedback and establishing the dominant mechanisms of energy transport from black holes to the ICM (e.g., sound waves and shocks vs turbulence, sloshing and/or 
buoyancy waves), and (ii) direct measurements of the composition of AGN bubbles inflated by
SMBH
in the centers of galaxy clusters with the aim in unravelling the impact of the direct cosmic ray (CR) heating on the ICM.\\

\vspace{-2mm} 
\noindent
{\large {\bf Physics of Feedback: Fluctuations in the Intracluster Medium}}\\

\vspace{-3mm} 
\indent
One plausible ICM heating mechanism 
is the thermalization of gas motions. These motions could be generated by propagating sound waves, rising bubbles of relativistic plasma, mergers, or galaxy motions. Their importance lies not only in their association with the heating processes, but also in their ability to reshape galaxies by propelling molecular outflows to galaxy-scale distances.
\begin{figure*}[!h]
\vspace{-0.2in}
\begin{center}
\includegraphics[width=0.9\textwidth]{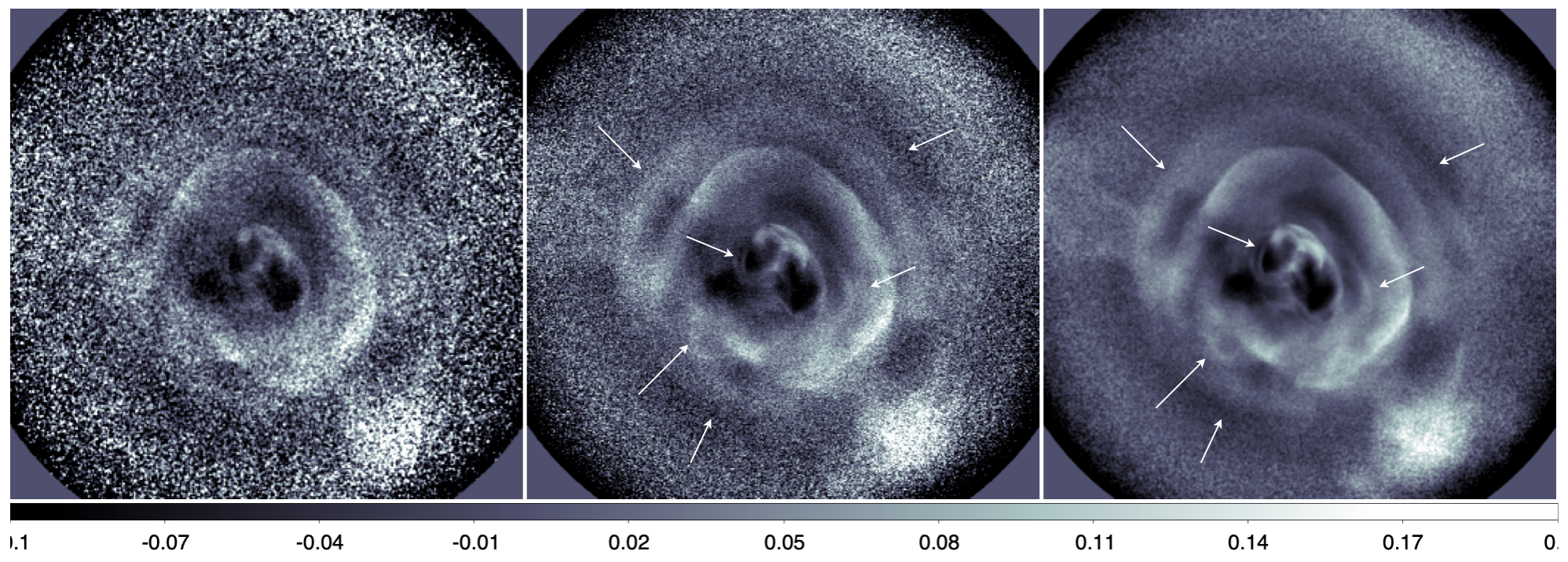}
\end{center}
\vspace{-0.3in}
\caption{\small Maps of adiabatic (sound waves) and isothermal (bubbles) fluctuations based on mock X-ray observations of a Perseus-like cluster. 
Photon counts in the soft and hard X-rays are 
processed such that isobaric fluctuations are removed \citep[see][for details]{Churazov2016}. 
\emph{Left:} 300ks  \chandra  observation 
\emph{Center:} 300ks {\it Lynx} observation 
\emph{Right:} 10Ms \chandra observation equivalent to a 1Ms observation with the next generation X-ray mission characterized by $\sim$10 times larger effective area. White arrows identify sharp structures associated with shocks, which will be detectable with future X-ray missions.}
\vspace{-10 pt}
\end{figure*}

\noindent
{\bf Black Hole--Generated Sound Waves, Sloshing Motions, and Turbulence} 
Supermassive black holes could deliver heat to the surrounding gas by generating sound waves and weak shocks. Coherent large-scale, approximately concentric ripples have been observed in the Perseus \cite{Fabian2003} and Virgo clusters \cite{Forman2005}, and other objects, and have been tentatively interpreted as sound waves. The dissipation of the mechanical energy of these waves could heat the ICM. This potential mode of heating is particularly appealing as the sound waves are expected to distribute the AGN energy very quickly (i.e., on the cool core sound crossing time) and in a quasi-isotropic fashion, thus offseting cooling while not violating the gas velocity dispersion constraints ($\sim$200 km s$^{-1}$) from the \hitomi mission \cite{Hitomi2018}. Furthermore, sound waves and weak shocks can carry a substantial fraction of the enthalpy contained in AGN cavities (e.g., \cite{McNamaraNulsen2007}). Interestingly, the fraction of AGN energy channeled into the waves should be systematically smaller for a given source power when the bubbles are dominated by CR pressure because, for a given bubble size, the bubble enthalpy is larger for a relativistic rather than thermal fluid. Direct measurements of AGN bubble expansion speeds with future high spectral and spatial resolution X-ray instruments could help to establish what fraction of the AGN energy is channeled to the rapidly propagating waves
(see Figure 2; \cite{Bruggen2005}).\\
\begin{wrapfigure}{R}{0.4 \textwidth}
\vspace{-15 pt}
\includegraphics[width = 0.4 \textwidth]{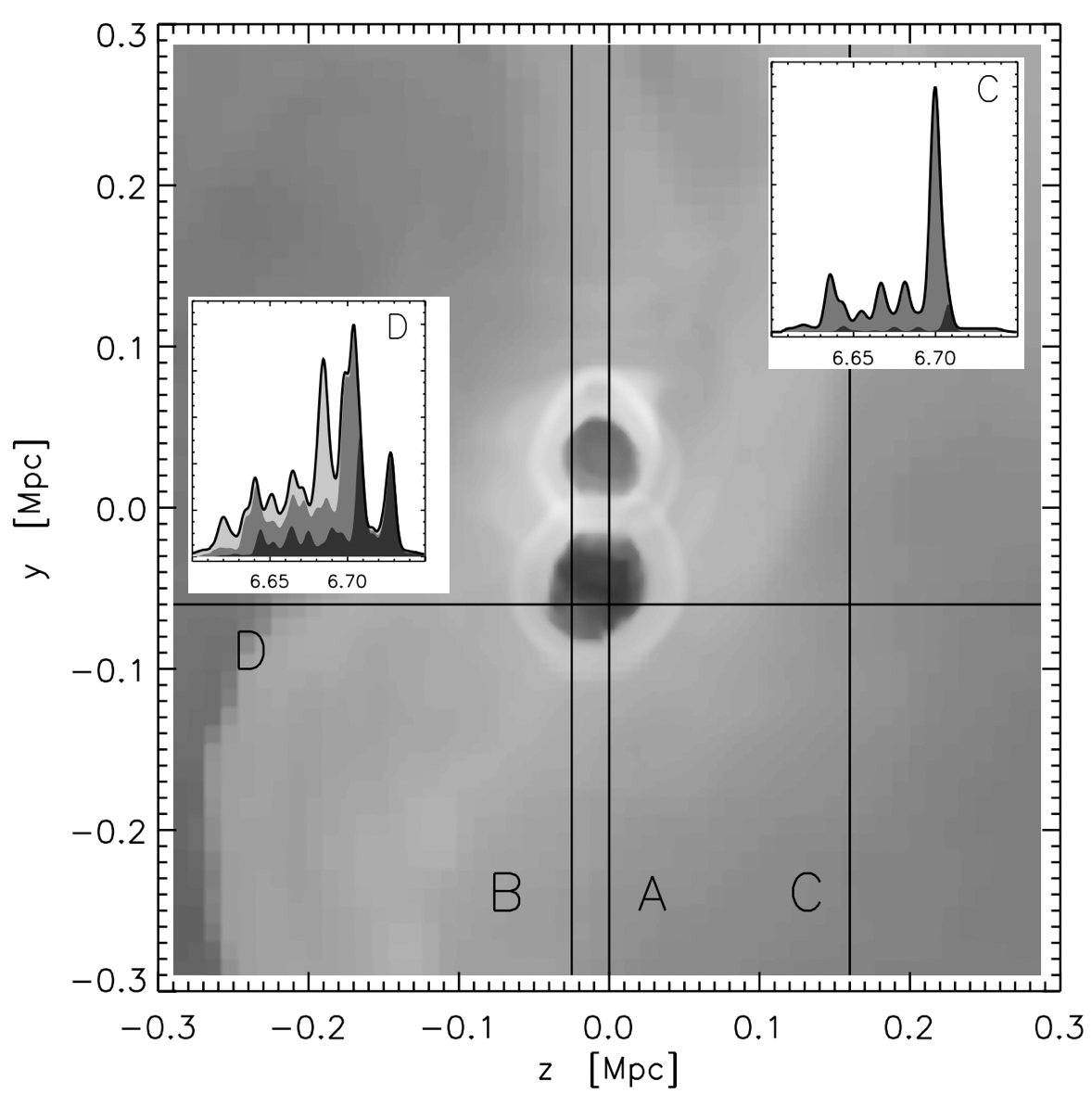}  
\vspace{-20 pt}
\caption{\small Simulation of AGN bubble inflation. Inlets show iron line profiles along the lines intersecting the bubble (D) and ambient medium (C). Line width $\sim$10 eV will be resolved by {\it Lynx}.; Adopted from \cite{Bruggen2005}.}
\vspace{-10 pt}
\end{wrapfigure}
\indent
Despite the promising features of this mode of heating, it is far from certain that the ripples are indeed sound waves. Coherent large-scale and approximately concentric fluctuations in the ICM could alternatively be caused by spiral sloshing gas motions \cite{ZuHone2013,Walker2018} 
(see also white paper by 
\cite{Markevitch2019}) 
or gravity waves and turbulence induced by the AGN in the gravitationally stratified ICM (e.g., \cite[][]{Chu00,Omm04,RuszkowskiOh2011, Zhu14b}). In the latter case, turbulence is expected to be weakly anisotropic with the preference for motions orthogonal to the radial direction.
By measuring velocities of gas motions indirectly through the analysis of X-ray surface brightness fluctuations (which are consistent with direct velocity measurements from {\it Hitomi}, see \cite{Hitomi2018, Zhuravleva2018}), it was shown that the dissipation of these motions provides enough heat to the ICM to offset radiative cooling (e.g., \cite{Zhu14b,Wal15}). 
In order to unravel the true nature of the fluctuations, and in particular to unambiguously determine the nature of the coherent ripples, the equation of state of the ICM fluctuations must be measured on small spatial scales. While the sound waves are adiabatic fluctuations ($dT/T = (2/3) d\rho/\rho$; where $T$ and $\rho$ are the temperature and density, respectively), the sloshing motions and turbulence generate predominantly isobaric ones ($dT/T = -d\rho/\rho$; e.g., \cite{Gaspari2014,Are16, Zhu16, Churazov2016}). Although current measurements show that fluctuations are mostly isobaric, the nature of fluctuations on small spatial scales dominated by ripple-like structures remains unknown due to a relatively poor spectral resolution and small effective area of current X-ray telescopes.\\
\indent
Figure 1 shows mock X-ray observations of a Perseus-like cluster processed using a method designed to filter out isobaric ICM fluctuations \cite{Churazov2016} based on hydrodynamical simulations of AGN feedback of \cite{Li2015}. 
Panels show 300 kilosecond {\it Chandra}, 300 kilosecond {\it Lynx}, and 10 million second \chandra mock observations.
While an exposure of 10 Ms is not within the reach of {\it Chandra}, collecting a comparable photon counts in a Ms observation may be within the grasp of future flagship X-ray missions characterized by high spatial resolution and $\sim$10 times larger effective area. 
This is a conservative estimate as the effective area is photon energy-dependent and may be even larger.  
This figure demonstrates that (adiabatic) sound waves could be unambiguously detected by future X-ray missions. Similar techniques could also be used to identify 
isobaric fluctuations, thus enabling one to 
\textcolor{NavyBlue}{\bf distinguish between contributions to the ICM heating from AGN feedback (via sound wave heating) and bulk gas 
motions (via dissipation of isobaric ICM fluctuations).}\\

\vspace{-3mm} 
\noindent
{\bf Microphysics of AGN Feedback} Detailed physics of AGN feedback on microphysical spatial scales is largely unknown, which 
limits our ability to reproduce realistic feedback in numerical simulations. Therefore,  one of the goals for the next decade is to unravel the nature of the ICM microphysics. 
Over the next decade, significant improvements in the capabilities of X-ray observatories are anticipated. The \xrism observatory, with the X-ray calorimeter, will be launched in early 2022. It will deliver spectra of extended X-ray sources with the resolution at least 30 times better than {\it Chandra}'s.
However, in order to understand transport processes in the ICM, it is necessary to study the statistical properties of gas density fluctuations and velocities on spatial scales that are comparable to the Coulomb mean free path (typically, $\sim$kpc to a few tens of kpc)  and below. On these scales, theoretical predictions for the behavior of the ICM vary substantially
depending on the properties of transport processes. The next-generation \lynx observatory would be the best mission for such science. It will have an effective area at least a factor of 10 larger than {\it Chandra}, a subarcsecond 
spatial resolution, and spectral resolution even better than {\it XRISM}. By measuring the \textcolor{NavyBlue}{\bf ICM fluctuation power spectra} in density, temperature, and velocity fluctuations, next generation missions may allow us to probe the mean-free-path scale and below
and quantify the level of thermal conduction and viscosity of the ICM (e.g., \cite{GaspariChurazov2013,Zhuravleva2018}).\\
\indent
Measurements of \textcolor{NavyBlue}{\bf sound wave} density, temperature, and velocity fluctuation amplitudes as a function of the distance from cluster centers could also help to \textcolor{NavyBlue}{\bf put constraints on the ICM transport processes}. Theoretical arguments suggest that ICM electron (and ion) conduction and ion viscosity should be significantly suppressed below their respective Spitzer-Braginskii values if the waves are to propagate out to significant distances from the AGN bubbles (e.g., \cite{Zweibel2018}) as appears to be the case if the ICM ripples are indeed sound waves. \\
\indent
Data from \xrism will also enable measurements of the gas motions generated by AGN and probe the velocity amplitudes, scales, anisotropy, and power spectra for a moderate sample of low-z clusters. It will enable studies of \textcolor{NavyBlue}{\bf non-equilibrium plasmas 
and put constraints of non-thermal electrons accelerated in AGN shocks} by measuring line ratios in high-resolution spectra \cite{Hitomi2018b}.\\

\vspace{-2mm} 
\noindent
{\large {\bf Physics of Feedback: Composition of AGN--inflated Bubbles}}\\

\vspace{-3mm} 
\indent
A major uncertainty in determining the power supplied by the AGN jets is associated with the fact that the composition of radio lobes inflated by the jets is unknown (e.g., \cite{DunnFabian2004,Birzan2008}). This poses a problem for 
future radio surveys that attempt to quantify the cosmological evolution of the energy injected by the 
supermassive black holes. 
It has recently been suggested that the composition of radio lobes in Fanaroff-Riley I and II radio galaxy populations (FR I and FR II, respectively) is systematically different. 
While FR I sources are likely dominated by hadronic CRs (mostly protons), FR II sources tend to be dominated by leptons (mostly electrons/positrons; \cite{Croston2018}). 
This implies that translating the jet radio power to its total power is very 
uncertain. For a given radio power, the total jet power can be orders of magnitude larger when the jets are heavier, i.e., hadron-dominated. 
Interestingly, recent multi-messenger neutrino and gamma-ray observations reported by the IceCube Collaboration \cite{IceCube2018} suggested that blazar jets can be sources of hadronic CRs.\\
\indent
Importantly, the amount of coupling between the energy contained in the AGN-inflated bubbles and the ambient ICM depends sensitively on the composition of the plasma filling them and on the physics of the CR transport processes (e.g., \cite{Ruszkowski2017,ProkhorovChurazov2017}). 
Consequently, the effective heating rates are expected to be different in the lepton- and hadron-dominated jet cases. Bubble composition will also have an impact on CR transport rates with dramatic consequences for the ICM dynamics and spatial uniformity of the distribution of AGN energy \cite{Ruszkowski2017,Yang2019}. Thus, it is imperative to unravel the composition of AGN cavities to understand how effectively the AGN energy is utilized in the ICM.\\
\indent
As discussed below, the composition of the cavities inflated by black holes has recently been probed more directly using the Sunyaev--Zel'dovich (SZ) effect (i.e., the inverse Compton (IC) scattering of CMB photons by the ICM). As we argue below, the composition of AGN outflows in clusters could be measured 
independently in X-rays using similar principles. 
Thus, 
the next generation X-ray and SZ instruments
may 
contribute to \textcolor{NavyBlue}{\bf (i) solving the long standing problem of the unknown composition of relativistic jets, and (ii) constraining the environment of the central AGN engine (jet launching, conversion of the black hole spin energy to the mechanical energy of outflows), thus ultimately linking the scales of gravitational radii 
($\sim$Solar System scales) to the scales of cool cores ($\sim$100 kpc)}.\\
\begin{figure*}[!h]
\vspace{-0.2in}
\begin{center}
\includegraphics[width=0.9\textwidth]{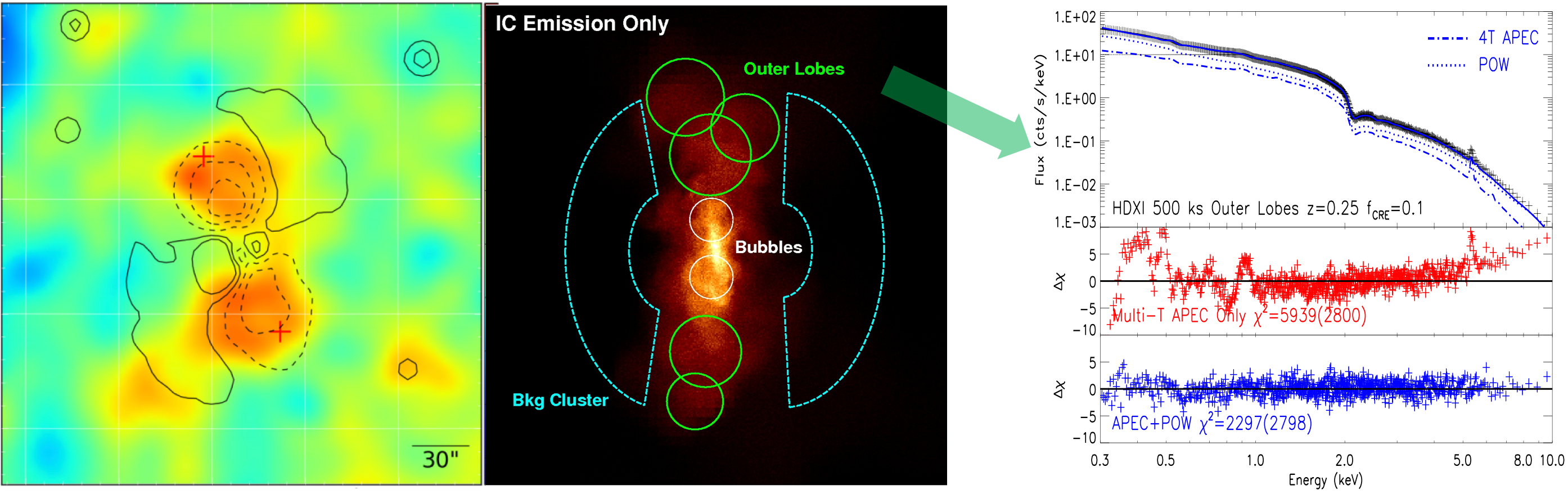}
\end{center}
\vspace{-0.2in}
\caption[]{
\small 
\emph{Left:} Residuals between the SZ 
the data and model (see text for details).
Positive residuals imply 
deficit
of the SZ signal, which is consistent with the picture of  CR-dominated AGN bubbles. 
\emph{Center:} Synthetic image of the inverse Compton X-ray signal from black hole inflated cavities in a galaxy cluster (assuming \lynx response)
\emph{Right:} AGN lobe spectra (top), multi-temperature 
fit residuals (middle) and the residuals corresponding to single temperature plus power law component representing IC emission (bottom). }
\vspace{-0.1in}
\end{figure*}

\vspace{-3mm} 
\noindent
{\bf Sunyaev--Zel'dovich Effect and the Composition of Black Hole--inflated Cavities} 
The SZ effect measures the integrated pressure along the line of sight in the ICM. While the interactions of thermal ICM electrons with the CMB lead to SZ signal, 
interactions with CRs contribute very little 
\cite{Pfrommer2005, Yang2019, Ehlert2019}. Recently, based on single-frequency Combined Array for Millimeter-wave Astronomy (\carma) observations, \cite{Abdulla2019} reported a 
suppression
of SZ signal from AGN cavities in the galaxy cluster MS 0735.6+7421. 
Figure 3 (left panel) shows 
the map of the residuals between the data and model based on the SZ signal corresponding to the smooth gas distribution inferred from X-ray observations 
and assuming pressure equilibrium between the bubbles and the ambient ICM. Positive residuals imply 
deficit
of the SZ signal, which
implies that the cavities may be 
either supported by very diffuse thermal plasma with temperature in excess of hundreds of keV, or are not supported thermally.
\textcolor{NavyBlue}{\bf Future high-angular resolution, multi-frequency SZ observations (CCAT-prime, NIKA2, MUSTANG2, TolTEC, AtLAST, LST, CSST, CMB-in-HD) can be expected to provide much sought-after proof of CR heating of the ICM} \cite[see][for more details]{Mroczkowski2019,Mroczkowski2019b,Sehgal2019}.\\

\vspace{-3mm} 
\noindent
{\bf X-ray Inverse Compton Effect Signal and AGN Bubble Composition} AGN cavities are often filled with radio-emitting plasma. The same electrons should be detectable in the hard X-rays (2-30 keV) due to IC scattering of the CMB. Accompanying these electrons may be relativistic protons that could contribute to, or even dominate, the pressure inside the bubbles to support them against the ambient ICM pressure. 
This theoretically expected IC signal could therefore constrain the particle content of black hole outflows and the long-term impact of these CRs on clusters. 
While in FR I 
sources 
this 
IC signal 
has so far eluded detection with \chandra or \xmm X-ray observatories, such measurements could be within the grasp of future X-ray missions combining large effective area and good spatial resolution, such as \lynx  or \axis \cite{Ruszkowski2019}. 
Figure 3 shows synthetic images (center panel) from a state-of-the-art magneto-hydrodynamical (MHD) simulation of black hole jet feedback that incorporates the effects of CR heating and transport. In these simulations, a small fraction of AGN energy is supplied in the form of CR leptons. Right panel shows spatially resolved spectra (top right panel), large multi-temperature thermal fit residuals (middle right panel), and much smaller residuals corresponding to single temperature gas plus power law component representing IC contribution to X-ray emission (lower right panel). This figure illustrates that this technique has the potential to rule out high-temperature thermal plasma filling the bubbles and can \textcolor{NavyBlue}{\bf directly detect non-thermal emission from CRs inside the AGN bubbles.} (Similar type of analysis for \chandra   instrument response demonstrated that it would be impossible to detect the IC signal.) This method is free from the assumptions of pressure equilibrium between the bubble and the ambient ICM, energy equipartition between the magnetic fields and CRs, and the characteristic bubble rise timescales often assumed to be comparable to the buoyancy timescale (e.g., \cite{Birzan2008, Croston2018}).

\pagebreak
\bibliographystyle{unsrturltrunc6}
\bibliography{refs}

\begin{thebibliography}{10}

\bibitem{Churazov2005}
E.~{Churazov}, S.~{Sazonov}, R.~{Sunyaev}, W.~{Forman}, C.~{Jones}, and
  H.~{B{\"o}hringer}.
\newblock {Supermassive black holes in elliptical galaxies: switching from very
  bright to very dim}.
\newblock {\em \mnras}, 363:L91--L95, October 2005.
\newblock \href {http://arxiv.org/abs/astro-ph/0507073}
  {\path{arXiv:astro-ph/0507073}}, \href
  {http://dx.doi.org/10.1111/j.1745-3933.2005.00093.x}
  {\path{doi:10.1111/j.1745-3933.2005.00093.x}}.

\bibitem{Fabian2012}
A.~C. {Fabian}.
\newblock {Observational Evidence of Active Galactic Nuclei Feedback}.
\newblock {\em \araa}, 50:455--489, September 2012.
\newblock \href {http://arxiv.org/abs/1204.4114} {\path{arXiv:1204.4114}},
  \href {http://dx.doi.org/10.1146/annurev-astro-081811-125521}
  {\path{doi:10.1146/annurev-astro-081811-125521}}.

\bibitem{Pillepich2018}
A.~{Pillepich}, V.~{Springel}, D.~{Nelson}, S.~{Genel}, J.~{Naiman},
  R.~{Pakmor}, et~al.
\newblock {Simulating galaxy formation with the IllustrisTNG model}.
\newblock {\em \mnras}, 473:4077--4106, January 2018.
\newblock \href {http://arxiv.org/abs/1703.02970} {\path{arXiv:1703.02970}},
  \href {http://dx.doi.org/10.1093/mnras/stx2656}
  {\path{doi:10.1093/mnras/stx2656}}.

\bibitem{Habouzit2019}
M.~{Habouzit}, S.~{Genel}, R.~S. {Somerville}, D.~{Kocevski}, M.~{Hirschmann},
  A.~{Dekel}, et~al.
\newblock {Linking galaxy structural properties and star formation activity to
  black hole activity with IllustrisTNG}.
\newblock {\em \mnras}, 484:4413--4443, April 2019.
\newblock \href {http://arxiv.org/abs/1809.05588} {\path{arXiv:1809.05588}},
  \href {http://dx.doi.org/10.1093/mnras/stz102}
  {\path{doi:10.1093/mnras/stz102}}.

\bibitem{Tremmel2019}
M.~{Tremmel}, T.~R. {Quinn}, A.~{Ricarte}, A.~{Babul}, U.~{Chadayammuri},
  P.~{Natarajan}, et~al.
\newblock {Introducing ROMULUSC: a cosmological simulation of a galaxy cluster
  with an unprecedented resolution}.
\newblock {\em \mnras}, 483:3336--3362, March 2019.
\newblock \href {http://arxiv.org/abs/1806.01282} {\path{arXiv:1806.01282}},
  \href {http://dx.doi.org/10.1093/mnras/sty3336}
  {\path{doi:10.1093/mnras/sty3336}}.

\bibitem{Werner2019}
N.~{Werner}, B.~R. {McNamara}, E.~{Churazov}, and E.~{Scannapieco}.
\newblock {Hot Atmospheres, Cold Gas, AGN Feedback and the Evolution of Early
  Type Galaxies: A Topical Perspective}.
\newblock {\em \ssr}, 215:5, January 2019.
\newblock \href {http://arxiv.org/abs/1811.05004} {\path{arXiv:1811.05004}},
  \href {http://dx.doi.org/10.1007/s11214-018-0571-9}
  {\path{doi:10.1007/s11214-018-0571-9}}.

\bibitem{Bolatto2019}
{A. Bolatto et al.}
\newblock {Cold Gas Outflows, Feedback, and the Shaping of Galaxies}.
\newblock {\em submitted to the Astro2020 Decadal Survey}.

\bibitem{Tremblay2019}
G.~R. {Tremblay}, E.~E. {Schneider}, A.~{Vikhlinin}, L.~{Hernquist},
  M.~{Ruszkowski}, B.~D. {Oppenheimer}, et~al.
\newblock {Galaxy Winds in the Age of Hyperdimensional Astrophysics}.
\newblock {\em arXiv e-prints}, March 2019.
\newblock \href {http://arxiv.org/abs/1903.05641} {\path{arXiv:1903.05641}}.

\bibitem{MathewsBrighenti2003}
W.~G. {Mathews} and F.~{Brighenti}.
\newblock {Hot Gas in and around Elliptical Galaxies}.
\newblock {\em \araa}, 41:191--239, 2003.
\newblock \href {http://arxiv.org/abs/astro-ph/0309553}
  {\path{arXiv:astro-ph/0309553}}, \href
  {http://dx.doi.org/10.1146/annurev.astro.41.090401.094542}
  {\path{doi:10.1146/annurev.astro.41.090401.094542}}.

\bibitem{Gaspari2012}
M.~{Gaspari}, M.~{Ruszkowski}, and P.~{Sharma}.
\newblock {Cause and Effect of Feedback: Multiphase Gas in Cluster Cores Heated
  by AGN Jets}.
\newblock {\em \apj}, 746:94, February 2012.
\newblock \href {http://arxiv.org/abs/1110.6063} {\path{arXiv:1110.6063}},
  \href {http://dx.doi.org/10.1088/0004-637X/746/1/94}
  {\path{doi:10.1088/0004-637X/746/1/94}}.

\bibitem{Churazov2016}
E.~{Churazov}, P.~{Arevalo}, W.~{Forman}, C.~{Jones}, A.~{Schekochihin},
  A.~{Vikhlinin}, et~al.
\newblock {Arithmetic with X-ray images of galaxy clusters: effective equation
  of state for small-scale perturbations in the ICM}.
\newblock {\em \mnras}, 463:1057--1067, November 2016.
\newblock \href {http://arxiv.org/abs/1605.08999} {\path{arXiv:1605.08999}},
  \href {http://dx.doi.org/10.1093/mnras/stw2044}
  {\path{doi:10.1093/mnras/stw2044}}.

\bibitem{Fabian2003}
A.~C. {Fabian}, J.~S. {Sanders}, S.~W. {Allen}, C.~S. {Crawford}, K.~{Iwasawa},
  R.~M. {Johnstone}, et~al.
\newblock {A deep Chandra observation of the Perseus cluster: shocks and
  ripples}.
\newblock {\em \mnras}, 344:L43--L47, September 2003.
\newblock \href {http://arxiv.org/abs/astro-ph/0306036}
  {\path{arXiv:astro-ph/0306036}}, \href
  {http://dx.doi.org/10.1046/j.1365-8711.2003.06902.x}
  {\path{doi:10.1046/j.1365-8711.2003.06902.x}}.

\bibitem{Forman2005}
W.~{Forman}, P.~{Nulsen}, S.~{Heinz}, F.~{Owen}, J.~{Eilek}, A.~{Vikhlinin},
  et~al.
\newblock {Reflections of Active Galactic Nucleus Outbursts in the Gaseous
  Atmosphere of M87}.
\newblock {\em \apj}, 635:894--906, December 2005.
\newblock \href {http://dx.doi.org/10.1086/429746} {\path{doi:10.1086/429746}}.

\bibitem{Hitomi2018}
{Hitomi Collaboration}, F.~{Aharonian}, H.~{Akamatsu}, F.~{Akimoto}, S.~W.
  {Allen}, L.~{Angelini}, et~al.
\newblock {Atmospheric gas dynamics in the Perseus cluster observed with
  Hitomi}.
\newblock {\em \pasj}, 70:9, March 2018.
\newblock \href {http://arxiv.org/abs/1711.00240} {\path{arXiv:1711.00240}},
  \href {http://dx.doi.org/10.1093/pasj/psx138}
  {\path{doi:10.1093/pasj/psx138}}.

\bibitem{McNamaraNulsen2007}
B.~R. {McNamara} and P.~E.~J. {Nulsen}.
\newblock {Heating Hot Atmospheres with Active Galactic Nuclei}.
\newblock {\em \araa}, 45:117--175, September 2007.
\newblock \href {http://arxiv.org/abs/0709.2152} {\path{arXiv:0709.2152}},
  \href {http://dx.doi.org/10.1146/annurev.astro.45.051806.110625}
  {\path{doi:10.1146/annurev.astro.45.051806.110625}}.

\bibitem{Bruggen2005}
M.~{Br{\"u}ggen}, M.~{Hoeft}, and M.~{Ruszkowski}.
\newblock {X-Ray Line Tomography of AGN-induced Motion in Clusters of
  Galaxies}.
\newblock {\em \apj}, 628:153--159, July 2005.
\newblock \href {http://arxiv.org/abs/astro-ph/0503656}
  {\path{arXiv:astro-ph/0503656}}, \href {http://dx.doi.org/10.1086/430732}
  {\path{doi:10.1086/430732}}.

\bibitem{ZuHone2013}
J.~A. {ZuHone}, M.~{Markevitch}, M.~{Ruszkowski}, and D.~{Lee}.
\newblock {Cold Fronts and Gas Sloshing in Galaxy Clusters with Anisotropic
  Thermal Conduction}.
\newblock {\em \apj}, 762:69, January 2013.
\newblock \href {http://arxiv.org/abs/1204.6005} {\path{arXiv:1204.6005}},
  \href {http://dx.doi.org/10.1088/0004-637X/762/2/69}
  {\path{doi:10.1088/0004-637X/762/2/69}}.

\bibitem{Walker2018}
S.~A. {Walker}, J.~{ZuHone}, A.~{Fabian}, and J.~{Sanders}.
\newblock {The split in the ancient cold front in the Perseus cluster}.
\newblock {\em Nature Astronomy}, 2:292--296, February 2018.
\newblock \href {http://arxiv.org/abs/1803.00898} {\path{arXiv:1803.00898}},
  \href {http://dx.doi.org/10.1038/s41550-018-0401-8}
  {\path{doi:10.1038/s41550-018-0401-8}}.

\bibitem{Markevitch2019}
M.~{Markevitch}, E.~{Bulbul}, E.~{Churazov}, S.~{Giacintucci}, R.~{Kraft},
  M.~{Kunz}, et~al.
\newblock {Physics of cosmic plasmas from high angular resolution X-ray imaging
  of galaxy clusters}.
\newblock {\em arXiv e-prints}, March 2019.
\newblock \href {http://arxiv.org/abs/1903.06356} {\path{arXiv:1903.06356}}.

\bibitem{Chu00}
E.~{Churazov}, W.~{Forman}, C.~{Jones}, and H.~{B{\"o}hringer}.
\newblock {Asymmetric, arc minute scale structures around NGC 1275}.
\newblock {\em \aap}, 356:788--794, April 2000.
\newblock \href {http://arxiv.org/abs/astro-ph/0002375}
  {\path{arXiv:astro-ph/0002375}}.

\bibitem{Omm04}
H.~{Omma}, J.~{Binney}, G.~{Bryan}, and A.~{Slyz}.
\newblock {Heating cooling flows with jets}.
\newblock {\em \mnras}, 348:1105--1119, March 2004.
\newblock \href {http://arxiv.org/abs/astro-ph/0307471}
  {\path{arXiv:astro-ph/0307471}}, \href
  {http://dx.doi.org/10.1111/j.1365-2966.2004.07382.x}
  {\path{doi:10.1111/j.1365-2966.2004.07382.x}}.

\bibitem{RuszkowskiOh2011}
M.~{Ruszkowski} and S.~P. {Oh}.
\newblock {Galaxy motions, turbulence and conduction in clusters of galaxies}.
\newblock {\em \mnras}, 414:1493--1507, June 2011.
\newblock \href {http://arxiv.org/abs/1008.5016} {\path{arXiv:1008.5016}},
  \href {http://dx.doi.org/10.1111/j.1365-2966.2011.18482.x}
  {\path{doi:10.1111/j.1365-2966.2011.18482.x}}.

\bibitem{Zhu14b}
I.~{Zhuravleva}, E.~{Churazov}, A.~A. {Schekochihin}, S.~W. {Allen},
  P.~{Ar{\'e}valo}, A.~C. {Fabian}, et~al.
\newblock {Turbulent heating in galaxy clusters brightest in X-rays}.
\newblock {\em \nat}, 515:85--87, November 2014.
\newblock \href {http://arxiv.org/abs/1410.6485} {\path{arXiv:1410.6485}},
  \href {http://dx.doi.org/10.1038/nature13830}
  {\path{doi:10.1038/nature13830}}.

\bibitem{Zhuravleva2018}
I.~{Zhuravleva}, S.~W. {Allen}, A.~{Mantz}, and N.~{Werner}.
\newblock {Gas Perturbations in the Cool Cores of Galaxy Clusters: Effective
  Equation of State, Velocity Power Spectra, and Turbulent Heating}.
\newblock {\em \apj}, 865:53, September 2018.
\newblock \href {http://arxiv.org/abs/1707.02304} {\path{arXiv:1707.02304}},
  \href {http://dx.doi.org/10.3847/1538-4357/aadae3}
  {\path{doi:10.3847/1538-4357/aadae3}}.

\bibitem{Wal15}
S.~A. {Walker}, J.~S. {Sanders}, and A.~C. {Fabian}.
\newblock {Constraining gas motions in the Centaurus cluster using X-ray
  surface brightness fluctuations and metal diffusion}.
\newblock {\em \mnras}, 453:3699--3705, November 2015.
\newblock \href {http://arxiv.org/abs/1508.04285} {\path{arXiv:1508.04285}},
  \href {http://dx.doi.org/10.1093/mnras/stv1929}
  {\path{doi:10.1093/mnras/stv1929}}.

\bibitem{Gaspari2014}
M.~{Gaspari}, E.~{Churazov}, D.~{Nagai}, E.~T. {Lau}, and I.~{Zhuravleva}.
\newblock {The relation between gas density and velocity power spectra in
  galaxy clusters: High-resolution hydrodynamic simulations and the role of
  conduction}.
\newblock {\em \aap}, 569:A67, September 2014.
\newblock \href {http://arxiv.org/abs/1404.5302} {\path{arXiv:1404.5302}},
  \href {http://dx.doi.org/10.1051/0004-6361/201424043}
  {\path{doi:10.1051/0004-6361/201424043}}.

\bibitem{Are16}
P.~{Ar{\'e}valo}, E.~{Churazov}, I.~{Zhuravleva}, W.~R. {Forman}, and
  C.~{Jones}.
\newblock {On the Nature of X-ray Surface Brightness Fluctuations in M87}.
\newblock {\em \apj}, 818:14, February 2016.
\newblock \href {http://arxiv.org/abs/1508.00013} {\path{arXiv:1508.00013}},
  \href {http://dx.doi.org/10.3847/0004-637X/818/1/14}
  {\path{doi:10.3847/0004-637X/818/1/14}}.

\bibitem{Zhu16}
I.~{Zhuravleva}, E.~{Churazov}, P.~{Ar{\'e}valo}, A.~A. {Schekochihin}, W.~R.
  {Forman}, S.~W. {Allen}, et~al.
\newblock {The nature and energetics of AGN-driven perturbations in the hot gas
  in the Perseus Cluster}.
\newblock {\em \mnras}, 458:2902--2915, May 2016.
\newblock \href {http://arxiv.org/abs/1601.02615} {\path{arXiv:1601.02615}},
  \href {http://dx.doi.org/10.1093/mnras/stw520}
  {\path{doi:10.1093/mnras/stw520}}.

\bibitem{Li2015}
Y.~{Li}, G.~L. {Bryan}, M.~{Ruszkowski}, G.~M. {Voit}, B.~W. {O'Shea}, and
  M.~{Donahue}.
\newblock {Cooling, AGN Feedback, and Star Formation in Simulated Cool-core
  Galaxy Clusters}.
\newblock {\em \apj}, 811:73, October 2015.
\newblock \href {http://arxiv.org/abs/1503.02660} {\path{arXiv:1503.02660}},
  \href {http://dx.doi.org/10.1088/0004-637X/811/2/73}
  {\path{doi:10.1088/0004-637X/811/2/73}}.

\bibitem{GaspariChurazov2013}
M.~{Gaspari} and E.~{Churazov}.
\newblock {Constraining turbulence and conduction in the hot ICM through
  density perturbations}.
\newblock {\em \aap}, 559:A78, November 2013.
\newblock \href {http://arxiv.org/abs/1307.4397} {\path{arXiv:1307.4397}},
  \href {http://dx.doi.org/10.1051/0004-6361/201322295}
  {\path{doi:10.1051/0004-6361/201322295}}.

\bibitem{Zweibel2018}
E.~G. {Zweibel}, V.~V. {Mirnov}, M.~{Ruszkowski}, C.~S. {Reynolds}, H.-Y.~K.
  {Yang}, and A.~C. {Fabian}.
\newblock {Acoustic Disturbances in Galaxy Clusters}.
\newblock {\em \apj}, 858:5, May 2018.
\newblock \href {http://arxiv.org/abs/1802.04808} {\path{arXiv:1802.04808}},
  \href {http://dx.doi.org/10.3847/1538-4357/aab9ae}
  {\path{doi:10.3847/1538-4357/aab9ae}}.

\bibitem{Hitomi2018b}
{Hitomi Collaboration}, F.~{Aharonian}, H.~{Akamatsu}, F.~{Akimoto}, S.~W.
  {Allen}, L.~{Angelini}, et~al.
\newblock {Atomic data and spectral modeling constraints from high-resolution
  X-ray observations of the Perseus cluster with Hitomi}.
\newblock {\em \pasj}, 70:12, March 2018.
\newblock \href {http://arxiv.org/abs/1712.05407} {\path{arXiv:1712.05407}},
  \href {http://dx.doi.org/10.1093/pasj/psx156}
  {\path{doi:10.1093/pasj/psx156}}.

\bibitem{DunnFabian2004}
R.~J.~H. {Dunn} and A.~C. {Fabian}.
\newblock {Particle energies and filling fractions of radio bubbles in cluster
  cores}.
\newblock {\em \mnras}, 355:862--873, December 2004.
\newblock \href {http://arxiv.org/abs/astro-ph/0409055}
  {\path{arXiv:astro-ph/0409055}}, \href
  {http://dx.doi.org/10.1111/j.1365-2966.2004.08365.x}
  {\path{doi:10.1111/j.1365-2966.2004.08365.x}}.

\bibitem{Birzan2008}
L.~{B{\^i}rzan}, B.~R. {McNamara}, P.~E.~J. {Nulsen}, C.~L. {Carilli}, and
  M.~W. {Wise}.
\newblock {Radiative Efficiency and Content of Extragalactic Radio Sources:
  Toward a Universal Scaling Relation between Jet Power and Radio Power}.
\newblock {\em \apj}, 686:859--880, October 2008.
\newblock \href {http://arxiv.org/abs/0806.1929} {\path{arXiv:0806.1929}},
  \href {http://dx.doi.org/10.1086/591416} {\path{doi:10.1086/591416}}.

\bibitem{Croston2018}
J.~H. {Croston}, J.~{Ineson}, and M.~J. {Hardcastle}.
\newblock {Particle content, radio-galaxy morphology, and jet power: all
  radio-loud AGN are not equal}.
\newblock {\em \mnras}, 476:1614--1623, May 2018.
\newblock \href {http://arxiv.org/abs/1801.10172} {\path{arXiv:1801.10172}},
  \href {http://dx.doi.org/10.1093/mnras/sty274}
  {\path{doi:10.1093/mnras/sty274}}.

\bibitem{IceCube2018}
{IceCube Collaboration}, M.~G. {Aartsen}, M.~{Ackermann}, J.~{Adams}, J.~A.
  {Aguilar}, M.~{Ahlers}, et~al.
\newblock {Multimessenger observations of a flaring blazar coincident with
  high-energy neutrino IceCube-170922A}.
\newblock {\em Science}, 361:eaat1378, July 2018.
\newblock \href {http://arxiv.org/abs/1807.08816} {\path{arXiv:1807.08816}},
  \href {http://dx.doi.org/10.1126/science.aat1378}
  {\path{doi:10.1126/science.aat1378}}.

\bibitem{Ruszkowski2017}
M.~{Ruszkowski}, H.-Y.~K. {Yang}, and C.~S. {Reynolds}.
\newblock {Cosmic-Ray Feedback Heating of the Intracluster Medium}.
\newblock {\em \apj}, 844:13, July 2017.
\newblock \href {http://arxiv.org/abs/1701.07441} {\path{arXiv:1701.07441}},
  \href {http://dx.doi.org/10.3847/1538-4357/aa79f8}
  {\path{doi:10.3847/1538-4357/aa79f8}}.

\bibitem{ProkhorovChurazov2017}
D.~A. {Prokhorov} and E.~M. {Churazov}.
\newblock {Confinement and diffusion time-scales of CR hadrons in AGN-inflated
  bubbles}.
\newblock {\em \mnras}, 470:3388--3394, September 2017.
\newblock \href {http://arxiv.org/abs/1707.05211} {\path{arXiv:1707.05211}},
  \href {http://dx.doi.org/10.1093/mnras/stx1404}
  {\path{doi:10.1093/mnras/stx1404}}.

\bibitem{Yang2019}
H.-Y.~K. {Yang}, M.~{Gaspari}, and C.~{Marlow}.
\newblock {The Impact of Radio AGN Bubble Composition on the Dynamics and
  Thermal Balance of the Intracluster Medium}.
\newblock {\em \apj}, 871:6, January 2019.
\newblock \href {http://arxiv.org/abs/1810.04173} {\path{arXiv:1810.04173}},
  \href {http://dx.doi.org/10.3847/1538-4357/aaf4bd}
  {\path{doi:10.3847/1538-4357/aaf4bd}}.

\bibitem{Pfrommer2005}
C.~{Pfrommer}, T.~A. {En{\ss}lin}, and C.~L. {Sarazin}.
\newblock {Unveiling the composition of radio plasma bubbles in galaxy clusters
  with the Sunyaev-Zel'dovich effect}.
\newblock {\em \aap}, 430:799--810, February 2005.
\newblock \href {http://arxiv.org/abs/astro-ph/0410012}
  {\path{arXiv:astro-ph/0410012}}, \href
  {http://dx.doi.org/10.1051/0004-6361:20041576}
  {\path{doi:10.1051/0004-6361:20041576}}.

\bibitem{Ehlert2019}
K.~{Ehlert}, C.~{Pfrommer}, R.~{Weinberger}, R.~{Pakmor}, and V.~{Springel}.
\newblock {The Sunyaev{\ndash}Zel{\rsquo}dovich Effect of Simulated
  Jet-inflated Bubbles in Clusters}.
\newblock {\em \apjl}, 872:L8, February 2019.
\newblock \href {http://arxiv.org/abs/1812.06982} {\path{arXiv:1812.06982}},
  \href {http://dx.doi.org/10.3847/2041-8213/ab020d}
  {\path{doi:10.3847/2041-8213/ab020d}}.

\bibitem{Abdulla2019}
Z.~{Abdulla}, J.~E. {Carlstrom}, A.~B. {Mantz}, D.~P. {Marrone}, C.~H. {Greer},
  J.~W. {Lamb}, et~al.
\newblock {Constraints on the Thermal Contents of the X-Ray Cavities of Cluster
  MS 0735.6+7421 with Sunyaev{\ndash}Zel{\rsquo}dovich Effect Observations}.
\newblock {\em \apj}, 871:195, February 2019.
\newblock \href {http://arxiv.org/abs/1806.05050} {\path{arXiv:1806.05050}},
  \href {http://dx.doi.org/10.3847/1538-4357/aaf888}
  {\path{doi:10.3847/1538-4357/aaf888}}.

\bibitem{Mroczkowski2019}
T.~{Mroczkowski}, D.~{Nagai}, K.~{Basu}, J.~{Chluba}, J.~{Sayers}, R.~{Adam},
  et~al.
\newblock {Astrophysics with the Spatially and Spectrally Resolved
  Sunyaev-Zeldovich Effects. A Millimetre/Submillimetre Probe of the Warm and
  Hot Universe}.
\newblock {\em \ssr}, 215:17, February 2019.
\newblock \href {http://arxiv.org/abs/1811.02310} {\path{arXiv:1811.02310}},
  \href {http://dx.doi.org/10.1007/s11214-019-0581-2}
  {\path{doi:10.1007/s11214-019-0581-2}}.

\bibitem{Mroczkowski2019b}
T.~{Mroczkowski}, D.~{Nagai}, P.~{Andreani}, M.~{Arnaud}, J.~{Bartlett},
  N.~{Battaglia}, et~al.
\newblock {A High-resolution SZ View of the Warm-Hot Universe}.
\newblock {\em arXiv e-prints}, March 2019.
\newblock \href {http://arxiv.org/abs/1903.02595} {\path{arXiv:1903.02595}}.

\bibitem{Sehgal2019}
N.~{Sehgal}, H.~N. {Nguyen}, J.~{Meyers}, M.~{Munchmeyer}, T.~{Mroczkowski},
  L.~{Di Mascolo}, et~al.
\newblock {Science from an Ultra-Deep, High-Resolution Millimeter-Wave Survey}.
\newblock {\em arXiv e-prints}, March 2019.
\newblock \href {http://arxiv.org/abs/1903.03263} {\path{arXiv:1903.03263}}.

\bibitem{Ruszkowski2019}
M.~{Ruszkowski}, E.~{Hodges-Kluck}, and H.-Y.~K. {Yang}.
\newblock {Unravelling cosmic ray composition of AGN bubbles with AXIS and
  Lynx}.
\newblock In {\em American Astronomical Society Meeting Abstracts \#233},
  volume 233 of {\em American Astronomical Society Meeting Abstracts}, page
  \#438.05, January 2019.

\end{thebibliography}

\end{document}